%% file: roper.tex
\documentclass[aps,twocolumn,prd,floatfix,tightenlines,final,preprintnumbers,
               superscriptaddress,showpacs,showkeys]{revtex4}
\usepackage{epsfig}
\usepackage{graphics}
\usepackage{latexsym}
\usepackage{amsmath}
\usepackage{amssymb}
\usepackage{rotating}
\usepackage{placeins}
\usepackage{hyperref}

\newcommand{\mpi}{m_\pi}
\begin{document}

\preprint{ADP-07-05/T645}
\preprint{JLAB-THY-07-642}
\preprint{Edinburgh 2007/9}

\title{Even parity excitations of the nucleon in lattice QCD}

\author{B.~G.~Lasscock}
\author{J.~N.~Hedditch}
\author{W.~Kamleh}
\author{D.~B.~Leinweber}
\affiliation{    Special Research Centre for the
                 Subatomic Structure of Matter,
                 and Department of Physics,
                 University of Adelaide, Adelaide SA 5005,
                 Australia}
\author{W.~Melnitchouk}
\affiliation{Jefferson Lab, 12000 Jefferson Avenue,
             Newport News, VA 23606, USA}
\author{A.~G.~Williams}
\affiliation{    Special Research Centre for the
                 Subatomic Structure of Matter,
                 and Department of Physics,
                 University of Adelaide, Adelaide SA 5005,
                 Australia}
\author{J.~M.~Zanotti}
\affiliation {School of Physics, University of  Edinburgh,
              Edinburgh EH9 3JZ, UK}

\begin{abstract}
  We study the spectrum of the even parity excitations of the nucleon
  in quenched lattice QCD.  We extend our earlier analysis by
  including an expanded basis of nucleon interpolating fields,
  increasing the physical size of the lattice, including more
  configurations to enhance statistics and probing closer to the
  chiral limit. With a review of world lattice data, we conclude that
  there is little evidence of the Roper resonance in quenched lattice
  QCD.
\end{abstract}

\vspace{3mm}

\maketitle

\section{Introduction}

One of the long-standing puzzles in baryon spectroscopy has been
the low mass of the first positive parity excitation of the nucleon,
the $J^P={1\over 2}^+$ Roper resonance, or $N^*(1440)$. 
In constituent (or valence) quark models with harmonic oscillator
quark-quark potentials, the lowest-lying odd parity ($J^P={1\over 2}^-$)
state naturally occurs below the positive parity radial excitation
(with principal quantum number $N=2$), whereas in nature the Roper
resonance is almost 100~MeV below the ${1\over 2}^-$ $N^*(1535)$ state.
Without fine tuning of parameters, valence quark models tend to
leave the mass of the Roper resonance too high.

Over the years various suggestions have been made to explain this
anomaly, ranging from speculations that the Roper resonance may be a
hybrid baryon state with excited glue \cite{Li:1991yb,Carlson:1991tg},
or a meson-baryon system \cite{Krehl:1999km}, or in terms of ``breathing
modes'' of the ground state nucleon \cite{Guichon:1985ny}.
To understand the nature of the Roper resonance in the context of QCD,
a number of studies have been performed recently within lattice QCD.

The study of excited baryons on the lattice has had a relatively
short history, although recently there has been growing interest
in identifying new techniques to isolate excited baryons, motivated
partly by the experimental $N^*$ program at Jefferson Lab.
The first detailed analysis of the positive parity excitation of the
nucleon was performed by Leinweber \cite{Leinweber:1994nm} using
Wilson fermions and an operator product expansion spectral ansatz.

In previous work by the CSSM Lattice Collaboration
\cite{Melnitchouk:2002eg} an analysis of the spectrum of octet baryons
was performed using the FLIC fermion action.
In each channel, a $2\times 2$ correlation matrix was used to extract
the low-lying states.
This approach was found to be successful in extracting the first excited
state of the negative parity $N^{*}(1535)$ state, and in the analysis of
the $\Lambda$ interpolating field.
However, the identification of the Roper resonance with this correlation
matrix remained elusive.

In the present study we extend the earlier work in several directions,
while focusing on the even parity nucleon spectrum.
In addition, we work with a larger lattice volume (2.5~fm compared to
1.95~fm in Refs.~\cite{Melnitchouk:2002eg,Zanotti:2003fx}), reducing
finite volume effects and enhancing the statistics.
Most importantly, we also use an expanded basis of interpolators compared
with that in Ref.~\cite{Melnitchouk:2002eg}, with the addition of the
spin-1/2 projected nucleon interpolator used in the calculation of the
spin-3/2 hadron mass spectrum \cite{Zanotti:2003fx}.

In the even parity spin-1/2 nucleon channel it is well known that the
two standard interpolating fields, which we label $\chi_{1}$ and
$\chi_{2}$, individually access the ground state and an excited state,
respectively.
The application of a $2\times 2$ correlation matrix with these
interpolators finds no evidence of a state with a mass different
from those that can already be extracted with the two interpolators
individually \cite{Melnitchouk:2002eg}.
Furthermore, the extracted excited state is found to be too massive to be
identified with the Roper resonance, and is therefore more likely to have
stronger overlap with the next even parity excited state of the nucleon
with mass 1710~MeV --- which we denote by $N'(1710)$ (in general we label
even parity nucleon excitations on the lattice by a superscript
$``\ '\ "$, and odd parity excitations on the lattice by a superscript
$``^{\ *}\ "$).
These findings are consistent with a similar correlation matrix analysis
by Sasaki et~al. \cite{Sasaki:2001nf}.
%

At the larger quark masses typically used in lattice calculations of
the spectrum, we expect that the three lowest-lying spin-1/2 even
parity states are the ground state nucleon, the Roper, and the second
even parity excited state (the $N'(1710)$), the latter which appears
to have strong coupling to the $\chi_2$ interpolator.  One would
therefore na\"ively expect that the addition of a third nucleon
interpolator to our basis should allow the mass of the Roper to be
extracted (in quenched lattice QCD).
In Ref.~\cite{Zanotti:2003fx} the spectra of the nucleon and $\Delta$
were analysed, including both spin-1/2 and spin-3/2 excited states.
For the nucleon spectrum a mixed spin-1/2, spin-3/2 interpolating field
(labeled $\chi_3$) was used.
The spin-1/2 projected $\chi_3$ interpolator was found to have good
overlap with the ground state, and in the present work we use this
interpolator as the third interpolating field.

In a similar analysis, Br\"ommel et~al. \cite{Brommel:2003jm} used the
$\chi_{1}$ and $\chi_{2}$ interpolators with the time-component of
the $\chi_3$ interpolator as a basis for a $3\times 3$ correlation
matrix analysis.
Even with the larger basis, Br\"ommel et~al. do not identify
the Roper on the lattice.
The difference between that study \cite{Brommel:2003jm} and our present
analysis is that we consider the spatial components of the $\chi_3$
interpolator, with spin-1/2 projection as in Ref.~\cite{Zanotti:2003fx}.

In Sec.~\ref{sec:existing} we review existing lattice calculations
of the positive parity excited nucleon spectrum and attempts to
identify the Roper resonance on the lattice.
Our lattice techniques are outlined in Sec.~\ref{sec:lattech}, where we
firstly summarise our simulation parameters and interpolating fields.
This is followed by a discussion of how to identify the spinor indices
in which the odd and even parity contributions to the correlation
functions propagate.
Our results are reported in Sec.~\ref{sec:res}, and conclusions
summarised in Sec.~\ref{sec:conc}.


\section{Existing lattice results}
\label{sec:existing}
\begin{figure*}[htbp]
\includegraphics[angle=90,height=21cm]{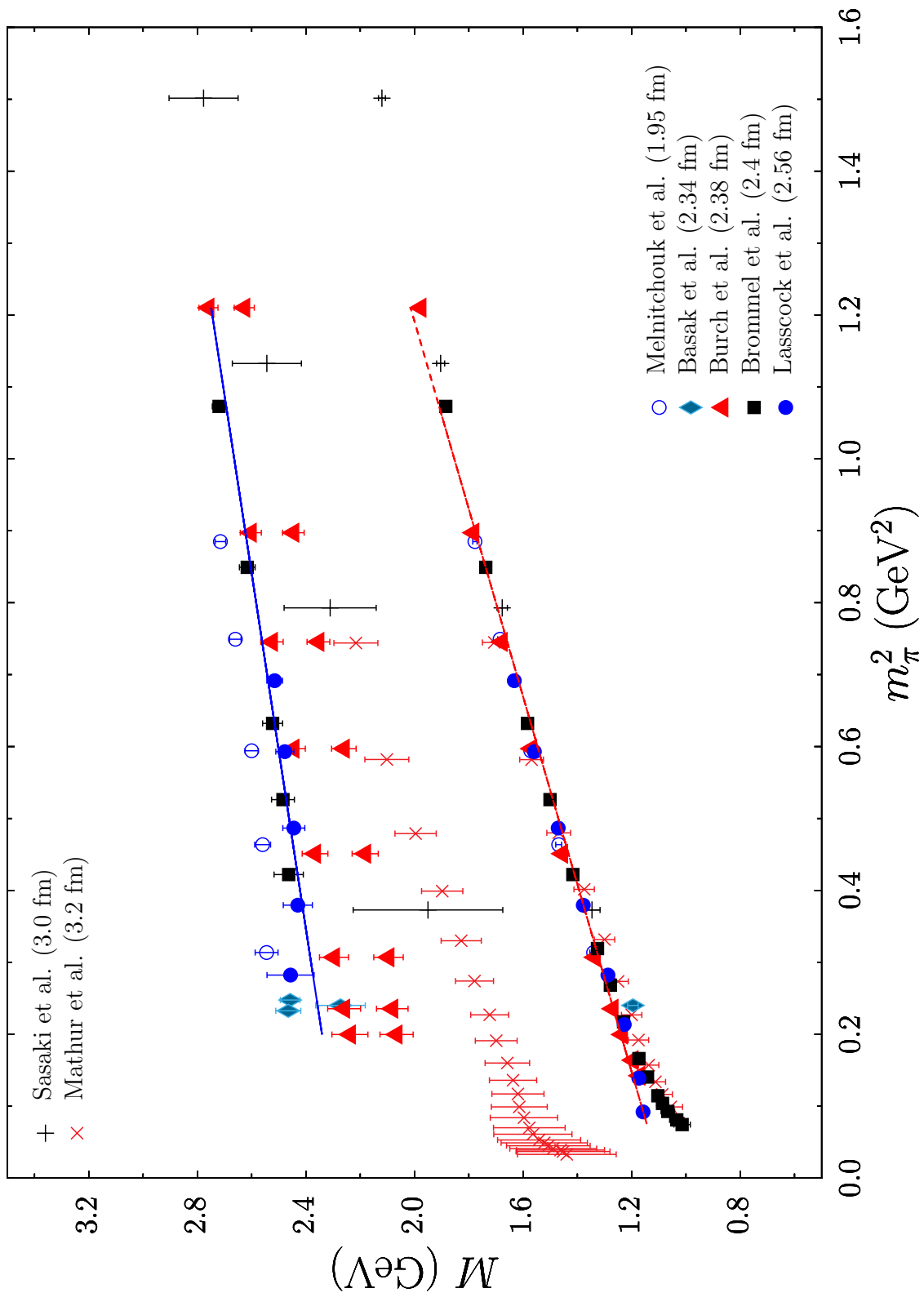}
\caption{\label{fig:compilation} Compilation of current lattice
  calculations of the spectrum of the even parity spin-1/2 excited
  states of the nucleon. In each study the ground state and excited
  state masses are shown.  Burch et~al.\ (closed triangles) report two
  excited states, one as the $N'$ the other as a Roper candidate.
  Basak et~al.\ (diamonds) report the masses of three excited states.
  Note that the two data points with degenerate masses from Basak
  et~al. \cite{Basak:2006ww} have been displaced horizontally for
  clarity. 
  To aid in understanding the results, lines of best fit to the $N'$ and 
  ground state masses are included.
  The solid line is a line of best fit to the $N'$ excited state
  masses extracted in this study, and the studies of Basak et~al.\ ,  Burch et~al.\
  and Brommel et~al.\ .  The dashed line is a line of best fit to
  the ground state mass for all of the data shown.}
\end{figure*}

In this section we review the findings of earlier lattice studies of
the spin-1/2, even parity nucleon mass spectrum.
Figure~\ref{fig:compilation} shows a compilation of recent
calculations of the mass spectrum in quenched lattice QCD.
Because the masses of the excited states at small lattice volumes are
expected to suffer from
significant finite volume effects, we focus only on those results obtained
on lattices with a physical size $\geq 2.0$~fm.

In Fig.~\ref{fig:compilation} the studies using Bayesian techniques
--- namely, Sasaki et~al. \cite{Sasaki:2005ap} (pluses) and Mathur
et~al.  \cite{Mathur:2003zf} (crosses) --- identify an excited state
which is interpreted as the Roper resonance.  The physical size of the
lattice in these analyses is 3.0 fm and 3.2~fm, respectively, and both
use point sources.  In both studies the mass of the odd parity excited
state is found to be consistent with that of the empirical
$N^{*}(1535)$ resonance.  At large quark masses, the level ordering of
the even and odd parity excited states is reversed compared to the
physical level ordering.  However, at small quark masses ($m_{\pi}$ as
small as 180~MeV) Mathur et~al.~\cite{Mathur:2003zf} find that the
correct empirical ordering appears to be restored.  However it is suggested by
Mathur et~al.\ that greater statistics are required to determine 
if this is true.

The masses in Fig.~\ref{fig:compilation} which are extracted
using a correlation matrix include the previous study by the CSSM
Lattice Collaboration \cite{Melnitchouk:2002eg} (open circles),
Br\"ommel et~al. \cite{Brommel:2003jm} (squares),
Burch et~al. \cite{Burch:2006cc} (triangles), and
Basak et~al. \cite{Basak:2006ww} (diamonds).
The mass of the excited state extracted with a $3\times 3$ correlation
matrix in the present analysis is also shown (filled circles).

The previous CSSM work \cite{Melnitchouk:2002eg} used the $\chi_1$ and
$\chi_2$ interpolators as a basis in the correlation, with a physical
lattice size of 1.92~fm.  At the fermion source 20 sweeps of gauge
invariant Gaussian smearing were used, with a smearing fraction of 0.7.
The findings strongly suggest that the mass of the extracted
excited state is too large to be identified with the Roper resonance.
These results sit somewhat high because of the relatively small volume
employed in the analysis.

A larger basis of operators is considered by Burch et~al. in
Ref.~\cite{Burch:2006cc}.
That study considers $\chi_{1}$, $\chi_{2}$ and an interpolator
equivalent to the temporal component of $\chi_{3}$ (up to an overall
factor of $\gamma_0$).
To expand the operator basis, two different fermion source and sink
Jacobi smearing prescriptions, labeled ``wide'' and ``narrow'',
are considered for each quark.
The narrow sources have 18 sweeps of smearing with a smearing
fraction of $\kappa = 0.210$, while the wide sources have 41 sweeps
with $\kappa = 0.191$.
The physical size of the lattice is 2.38~fm.
In the analysis of Ref.~\cite{Burch:2006cc} the basis of operators is
restricted to $\chi_{1}$ and the $\chi_{3}$-like interpolator with
three different smearing prescriptions for each interpolator,
making a total of six different operators.  

In Refs.~\cite{Burch:2006cc} and \cite{Burch:2004he} Burch et~al.
argue that in the limit of large quark mass, the state corresponding
to the $N'$ can be identified by comparison with the mass
extracted with the $\chi_{2}$ interpolator.
They consequently identify the largest mass state with the $N'$
and conclude that the lower energy state is therefore the Roper.
We note that at the larger quark masses the lower energy state,
identified as the Roper with the Bayesian techniques in 
Ref.~\cite{Mathur:2003zf}, is similar to the mass of the same
state identified using a correlation matrix analysis in 
Ref.~\cite{Burch:2006cc}.
However, the two techniques disagree at the smaller quark masses.

Basak et~al. \cite{Basak:2006ww} take advantage of the
discrete symmetries on the lattice and identify a large basis of local
and non-local operators \cite{Basak:2005ir}.
The complete set of local and singly displaced non-local sources are
used as a basis of their correlation matrix, and two distinct excited
states are identified.
The interpretation of Basak et~al. \cite{Basak:2006ww} is that the lower
energy excited state corresponds to the $N'$.
This state is consistent with the higher energy state found by Burch
et~al. \cite{Burch:2006cc}, and our new results presented herein.


The best fit to the excited state data extracted on a lattice with
physical size $\sim 2.5$~fm using the correlation matrix technique
is shown by the solid line in Fig.~\ref{fig:compilation}.
We fit the data for the largest energy state extracted by Burch et~al.
\cite{Burch:2006cc}, along with the data from Basak et~al.
\cite{Basak:2006ww} and Br\"ommel et~al. \cite{Brommel:2003jm},
and the results of the present study.
Our results are consistent with the masses extracted
by Br\"ommel et~al.\ and by Burch et~al. at their largest quark masses.
At the smaller quark masses, the highest mass states obtained by Burch
et~al.\ and Basak et~al.\ lie on either side of the line of best fit.
Since these two analyses use different gauge and quark
actions, it is not possible to determine whether the small discrepancy
between their masses is of any statistical significance.
It is also likely that the operators considered in each study have
different couplings to the $N+\pi$ in P-wave (and equivalently the
P-wave $N+\eta'$ in quenched QCD) scattering state.
We know that as smaller quark masses are approached, the level ordering
between the lowest energy multi-hadron state and the $N'$ state is
reversed on the lattice.


\section{ Lattice Techniques }
\label{sec:lattech}

The present analysis is based on an ensemble of 396 gauge-field
configurations on a $20^{3}\times 40$ lattice, using the mean-field
${\cal O}(a^2)$-improved Luscher-Weisz plaquette plus rectangle gauge
action \cite{Luscher:1984xn}.
The lattice spacing is 0.128~fm, set with the Sommer scale 
$r_{0}=0.49$~fm.
For the fermion propagators we use the FLIC fermion action 
\cite{Zanotti:2001yb}, which is an ${\cal O}(a)$-improved action with
excellent scaling properties, providing near continuum results at finite
lattice spacing \cite{Zanotti:2004dr}.
A fixed boundary condition in the time direction is implemented by
setting $U_t(\vec x, N_t) = 0\ \forall\ \vec x$ in the hopping terms
of the fermion action.
Periodic boundary conditions are imposed in the spatial directions.
We find that the fixed boundary effects are only significant after
time slice 30 \cite{Lasscock:2005kx}, which is the limit of our
analysis of the correlation functions presented below.

We apply $36$ sweeps of gauge-invariant Gaussian smearing
\cite{Gusken:1989qx}, with smearing fraction $\alpha = 0.7$, in the
spatial dimensions at the fermion source ($t=8$).  Eight quark masses
are considered in this calculation, providing $a m_\pi = \{$ $0.540$,
$0.500$, $0.453$, $0.400$, $0.345$, $0.300$, $0.242$, $0.197$ $\}$.
The error analysis is performed by a second-order, single-elimination
jackknife, with the $\chi^2$ per degree of freedom obtained via
covariance matrix fits.  Further details of the fermion action and
simulation parameters can be found in
Refs.~\cite{Zanotti:2001yb,Zanotti:2004dr} and
\cite{Lasscock:2005tt,Lasscock:2005kx}, respectively.
We apply the variational method as discussed in
Refs.~\cite{Melnitchouk:2002eg,Lasscock:2005tt}.

\subsection{ Interpolating Fields }
\label{ssec:ints}

In the previous CSSM study \cite{Melnitchouk:2002eg} the $\chi_1$ and
$\chi_2$ interpolators were found not to have significant overlap
with each other.
Following the approach by Br\"ommel et~al. \cite{Brommel:2003jm}, we
extend this analysis by including the $\chi_3$ interpolator that was
used by Zanotti et~al. \cite{Zanotti:2003fx} to extract spin-3/2 nucleon
excited states. 
Our basis of nucleon interpolating fields is then given by:
\begin{eqnarray}
\label{eq:ints}
\chi_{1}(x)       &=& \epsilon^{abc} (u^{T a}(x) C \gamma_5  d^b(x)) u^{c}(x)\ , \cr
\chi_{2}(x)       &=& \epsilon^{abc} (u^{T a}(x) C           d^b(x)) \gamma_{5} u^{c}(x)\ , \cr
\chi^{\mu}_{3}(x) &=& \epsilon^{abc} (u^{T a}(x) C \gamma_5 \gamma^{\mu} d^b(x)) \gamma_{5} u^{c}(x)\ .
\end{eqnarray}
In all of our phenomenology we use the Dirac representation of the
$\gamma$-matrices. 

\subsection{ Excited baryons on the lattice }
\label{sec:excited_baryons}

We begin our discussion with a review of how the masses of even and odd
parity states are extracted from the correlation function using the
spin-1/2 $\chi_1$ and $\chi_2$ interpolators.
On the baryon level, the two-point correlation function in momentum
space is:
\begin{eqnarray}
\label{eq:2-pt}
\mathcal{G}(t,{\vec p}) &=& \sum_{\vec x}\ e^{-i {\vec p} \cdot {\vec x}}
\left\langle 0 \left|
   T\ \chi(x)\ \bar\chi(0)\
\right| 0 \right\rangle\ ,
\end{eqnarray}
where the interpolator $\chi (\bar{\chi})$ annihilates (creates) baryon
states to (from) the vacuum. 
Inserting a complete set of intermediate momentum, energy and spin
states $| B,\vec{p'},s \rangle$,
\begin{eqnarray}
\label{eq:completeness}
1 = \sum_{B,\vec{p}\hspace{1mm}',s} | B,\vec{p}\hspace{1mm}',s \rangle\langle B,\vec{p}\hspace{1mm}',s |\ ,
\end{eqnarray}
we obtain
\begin{widetext}
\begin{eqnarray}
\mathcal{G}(t,{\vec p})
= { \sum_{s,\vec{p}\hspace{1mm}',B} }\sum_{\vec x} e^{-i {\vec p} \cdot {\vec x}} \
  \langle 0 |\ \chi(x)\ | B,\vec{p}\hspace{1mm}',s \rangle
  \langle B,\vec{p}\hspace{1mm}',s |\ \bar\chi(0)\ | 0 \rangle\ ,
\end{eqnarray}
where the state $B$ has mass $M_B$ and energy
$E_B = \sqrt{M_B^2 + \vec p^2}$.
The sum over all possible states $B$ with a given set of quantum numbers
includes a tower of resonances and multi-hadron states created by our
interpolators.
Using 
$\chi(x)=e^{iP\cdot x} \chi(0) e^{-iP\cdot x}$,
where $P$ is the four-momentum operator, we can write:
\begin{eqnarray}
\label{eq:2-pt.master}
\mathcal{G}(t,{\vec p})
&=& { \sum_{s,\vec{p}\hspace{1mm}',B} }\sum_{\vec x} e^{-i {\vec p} \cdot {\vec x}} \
  \langle 0 |\ e^{iP\cdot x} \chi(0)  e^{-iP\cdot x}\ | B,\vec{p}\hspace{1mm}',s \rangle
  \langle B,\vec{p}\hspace{1mm}',s |\ \bar\chi(0)\ | 0 \rangle\cr
&=& { \sum_{s,\vec{p}\hspace{1mm}',B} } e^{-iE_{B}t}\sum_{\vec x}e^{-i \vec{x} \cdot( {\vec p} - \vec{p}\hspace{1mm}') }\
  \langle 0 |\ \chi(0)\ | B,\vec{p}\hspace{1mm}',s \rangle
  \langle B,\vec{p}\hspace{1mm}',s |\ \bar\chi(0)\ | 0 \rangle \cr
&=& { \sum_{s,\vec{p}\hspace{1mm}',B} } e^{-iE_{B}t} \delta_{\vec{p}\vec{p}\hspace{0.5mm}'}\
  \langle 0 |\ \chi(0)\ | B,\vec{p}\hspace{1mm}',s \rangle
  \langle B,\vec{p}\hspace{1mm}',s |\ \bar\chi(0)\ | 0 \rangle\cr
&\to& { \sum_{B} } e^{-E_{B}t} { \sum_{s} }\
  \langle 0 |\ \chi(0)\ | B,\vec{p},s \rangle
  \langle B,\vec{p},s |\ \bar\chi(0)\ | 0 \rangle \ ,
\end{eqnarray}
where on the last line we make the replacement $it \to t$ for
Euclidean time.

Next we evaluate the matrix elements in Eq.~(\ref{eq:2-pt.master}),
labeling the even and odd parity contributions to the correlation
function by ``$+$'' and ``$-$'', respectively.
The overlap of $\chi$ and $\bar{\chi}$ with even and odd parity
baryons, such as the nucleon for example, can be expressed as:
\begin{eqnarray}
\label{eq:even_matrix_element}
\langle 0 |\ \chi(0)\ | N_{1/2^{+}}(\vec{p},s) \rangle     &=& \lambda_{N_{1/2^{+}}}\ \sqrt{\frac{M_{N_{1/2^{+}}}}{E_{N_{1/2^{+}}}}} u(p_{N_{1/2^{+}}},s) \ , \cr
\langle 0 |\ \chi(0)\ | N_{1/2^{-}}(\vec{p},s) \rangle     &=& \lambda_{N_{1/2^{-}}}\ \sqrt{\frac{M_{N_{1/2^{-}}}}{E_{N_{1/2^{-}}}}} \gamma_{5}u(p_{N_{1/2^{-}}},s) \ , \cr
\langle N_{1/2^{+}}(\vec{p},s) |\ \bar\chi(0)\ | 0 \rangle &=& \bar{\lambda}_{N_{1/2^{+}}}\ \sqrt{\frac{M_{N_{1/2^{+}}}}{E_{N_{1/2^{+}}}}} \bar{u}(p_{N_{1/2^{+}}},s) \ , \cr
\langle N_{1/2^{-}}(\vec{p},s) |\ \bar\chi(0)\ | 0 \rangle &=& - \bar{\lambda}_{N_{1/2^{-}}}\ \sqrt{\frac{M_{N_{1/2^{-}}}}{E_{N_{1/2^{-}}}}} \bar{u}(p_{N_{1/2^{-}}},s)\gamma_{5} \ ,
\end{eqnarray}
where $u(p,s)$ is a Dirac spinor and $\lambda (\bar{\lambda})$ are
couplings of the interpolators at the sink (source).
Note that the four-momentum $p_{N_{1/2^{+}}}$ is on-shell, with
$p_{0}=\sqrt{\vec{p}^{2} + M^{2}_{N_{1/2^{+}}}}$.
Because the fermion source is smeared, the coupling $\bar{\lambda}$
is not equal to the adjoint of $\lambda$.

\end{widetext}

Substituting the appropriate terms and using the identity
\begin{eqnarray}
\label{eq:spin_sum}
\sum_{s} u(p,s)\bar{u}(p,s) = { ( \gamma \cdot p + M ) \over 2M }\ ,
\end{eqnarray}
the contributions of the even and odd parity terms to the correlation
function can be written as:
\begin{eqnarray}
\mathcal{G}(t,{\vec p})
&=& { \sum_{B^{+}} } \lambda_{B^{+}}\bar{\lambda}_{B^{+}}e^{-E_{B^{+}}t}{ ( \gamma \cdot p_{B^{+}} + M_{B^{+}} ) \over 2E_{B^{+}} } \cr
&+& { \sum_{B^{-}} } \lambda_{B^{-}}\bar{\lambda}_{B^{-}}e^{-E_{B^{-}}t}{ ( \gamma \cdot p_{B^{-}} - M_{B^{-}} ) \over 2E_{B^{-}} } \ . 
\end{eqnarray}
The masses of states with definite parity can then be obtained
from the spinor trace of the parity projected correlation functions,
\begin{eqnarray}
G_{\pm}(t,{\vec 0})
&=& {\rm tr_{sp}}
    \left[ \Gamma_{\pm} \, \mathcal{G}(t,{\vec 0})\, \right]						\cr
&=& \sum_{B^{\pm}} \lambda_{B^{\pm}} \bar{\lambda}_{B^{\pm}}
    \exp{(-E_{B^{\pm}} t)}                                      \cr
&\stackrel{t\to\infty}{=}& \lambda_{0^{\pm}} \bar{\lambda}_{0^{\pm}} \exp{(-M_{0^\pm} t)}\ ,
\end{eqnarray}
%
%
where $\Gamma_{\pm} = {\gamma_{0} \pm 1 \over 2}$ is the parity
projection operator at zero momentum \cite{Lee:1998cx},
and the subscripts $0^{\pm}$ label the lowest energy state with the
projected quantum numbers. 

\begin{widetext}
For the $\chi_{3}^\mu$ interpolator the analogue of
Eq.~(\ref{eq:2-pt.master}) is given by:
\begin{eqnarray}
\label{eq:2-pt.master.spin32}
\mathcal{G}^{\mu\nu}(t,{\vec p}) &=& { \sum_{B} } e^{-E_{B}t}  \sum_{s}\
  \langle 0 |\ \chi_{3}^{\mu}(0)\ | B,\vec{p},s \rangle
  \langle B,\vec{p},s |\ \bar\chi_{3}^{\nu}(0)\ | 0 \rangle \ ,
\end{eqnarray}
where now $\chi_{3}^\mu$ overlaps with both spin-1/2 and spin-3/2 states.
In this study we project spin-1/2 states using the spin projection
operator discussed in Ref.~\cite{Zanotti:2003fx},
\begin{eqnarray}
P^{\frac{1}{2}}_{\mu\nu}(p)
&=&  \frac{1}{3}\gamma_{\mu}\gamma_{\nu}
 + \frac{1}{3p^{2}}(\gamma\cdot p \gamma_{\mu}p_{\nu}
		   + p_{\mu}\gamma_{\nu}\gamma\cdot p)\ .
\end{eqnarray}
At zero momentum $p = (M,0,0,0)$, relevant to the mass determination, the 
spin-projection operator has no hypercubic lattice artifacts.

We proceed in our analysis of the $\chi_3^\mu$ interpolator by evaluating
the analogue of the matrix element in Eqs.~\eqref{eq:even_matrix_element}.
Following Zanotti et~al. \cite{Zanotti:2003fx}, the coefficient of the
spinor is taken to be a linear combination of four-vectors.
The requirement that the matrix elements transform as pseudovectors under
parity restricts the coefficients to be proportional to either the
four-momentum or the matrix $\gamma^{\mu}$.
The analogue of Eqs.~\eqref{eq:even_matrix_element} can therefore be
written as:
\begin{eqnarray}
\label{eq:lorentz_pseudovector_elements}
\hspace{-2mm}\langle 0 | \chi^{\mu}_3(0) | N_{1/2^{+}}(\vec{p},s) \rangle \hspace{-2mm} &=& \hspace{-2mm}
(\alpha_{N_{1/2^+}}p_{N_{1/2^+}}^{\mu} \hspace{-1mm} + \hspace{-1mm} \beta_{N_{1/2^+}}\gamma^{\mu})
\sqrt{ {M_{N_{1/2^+}} \over E_{N_{1/2^+}}} }\ \gamma_5 u(p_{N_{1/2^{+}}},s), \\
\hspace{-2mm}\langle 0 | \chi^{\mu}_3(0) | N_{1/2^{-}}(\vec{p},s) \rangle \hspace{-2mm} &=& \hspace{-2mm}
(\alpha_{N_{1/2^-}}p_{N_{1/2^-}}^{\mu} \hspace{-1mm} + \hspace{-1mm} \beta_{N_{1/2^-}}\gamma^{\mu})
\sqrt{ {M_{N_{1/2^-}} \over E_{N_{1/2^-}}} }\ u(p_{N_{1/2^-}},s), \\
\hspace{-2mm}\langle N_{1/2^{+}}(\vec{p},s) | \bar{\chi}^{\mu}_3(0) | 0 \rangle \hspace{-2mm} &=& \hspace{-2mm} 
- \sqrt{ {M_{N_{1/2^+}} \over E_{N_{1/2^+}}} }\ \bar{u}(p_{N_{1/2^+}},s) \gamma_5
(\bar{\alpha}_{N_{1/2^+}}p_{N_{1/2^+}}^{\mu} \hspace{-1mm} + \hspace{-1mm} \bar{\beta}_{N_{1/2^+}}\gamma^{\mu}), \\
\hspace{-2mm}\langle N_{1/2^{-}}(\vec{p},s) | \bar{\chi}^{\mu}_3(0) | 0 \rangle \hspace{-2mm} &=& \hspace{-2mm}
 \sqrt{ {M_{N_{1/2^-}} \over E_{N_{1/2^-}}} }\ \bar{u}(p_{N_{1/2^-}},s)
(\bar{\alpha}_{N_{1/2^-}}p_{N_{1/2^-}}^{\mu} \hspace{-1mm} + \hspace{-1mm}\bar{\beta}_{N_{1/2^-}} \gamma^{\mu}),
\end{eqnarray}
where the factors $\alpha_B$ and $\beta_B$ denote the coupling
strengths of the interpolating field $\chi^\mu_3$ to the baryon $B$,
and similarly for the adjoint.
Combining these expressions with their respective adjoints, and using
Eq.~\eqref{eq:spin_sum} for the energy projector,
the contribution to the correlation function from spin-1/2 states
extracted with $\chi_{3}^{\mu}$ is:
\begin{eqnarray}
\mathcal{G}(t,{\vec p}) &=& { \sum_{B^{+}} } e^{-E_{B^{+}}t}\
 (\alpha_{N_{1/2^+}}p_{N_{1/2^+}}^{\nu} + \beta_{N_{1/2^+}}\gamma^{\nu})\
\gamma_5 \frac{\gamma\cdot p_{N_{1/2^+}} + M_{N_{1/2^+}}}{2E_{N_{1/2^+}}} \gamma_5
(\bar{\alpha}_{N_{1/2^+}}p_{N_{1/2^+}}^{\nu} + \bar{\beta}_{N_{1/2^+}}
\gamma^{\nu})\nonumber  \\
& & \hspace{-0.5cm} +\
{ \sum_{B^{-}} } e^{-E_{B^{-}}t} \
 (\alpha_{N_{1/2^-}}p_{N_{1/2^-}}^{\nu} + \beta_{N_{1/2^-}}\gamma^{\nu})\
\frac{\gamma\cdot p_{N_{1/2^-}} + M_{N_{1/2^-}}}{2E_{N_{1/2^-}}} 
(\bar{\alpha}_{N_{1/2^-}}p_{N_{1/2^-}}^{\nu} + \bar{\beta}_{N_{1/2^-}}
\gamma^{\nu})\ .
\end{eqnarray}
In this analysis we average over the contributions to the
spin-projected correlation function with $\mu=1$--3 and $\nu=1$--3.
We need not evaluate the $(\mu,\nu)=(k,0)$ or $(0,k)$ terms, where $k=1$--3, as
these do not contribute to the correlation function after spin
projection at $\vec{p} = \vec{0}$.

Finally, we evaluate the two-point function at the hadronic level for
the cross correlators
$\langle 0 | T \chi^{\mu}_3\bar{\chi}_{i} | 0 \rangle$ and
$\langle 0 | T \chi_{i}\bar{\chi}^{\mu}_{3} | 0 \rangle$,
with $i=1,2$.
It is important to note that as these correlation functions are not
Lorentz scalars they remain dependent on the representation of the
$\gamma$-matrices.
The respective correlation functions are given by:
\begin{eqnarray}
\label{eq:2-pt.cmatrix-mux}
\mathcal{G}_{3i}^{\mu}(t,{\vec p})
&=& \sum_{\vec x}\ \exp({-i {\vec p} \cdot {\vec x}}) 
\left\langle 0 \left|
   T\ \chi_{3}^{\mu}(x)\ \bar\chi_{i}(0)\
\right| 0 \right\rangle\ ,	\\
\label{eq:2-pt.cmatrix-xmu}
\mathcal{G}_{i3}^{\mu}(t,{\vec p})
&=& \sum_{\vec x}\ \exp({-i {\vec p} \cdot {\vec x}}) 
\left\langle 0 \left|
   T\ \chi_{i}(x)\ \bar\chi_{3}^{\mu}(0)\
\right| 0 \right\rangle\ .
\end{eqnarray}
As for the diagonal correlators discussed above, we proceed by inserting
a complete set of states and evaluating the resulting matrix elements.  
For the function $\mathcal{G}_{i3}^{\mu}$, we can use
Eqs.~\eqref{eq:even_matrix_element}, \eqref{eq:spin_sum} and
\eqref{eq:lorentz_pseudovector_elements} to write the matrix
elements as:
\begin{eqnarray}
& &\sum_{s} \langle 0 | \chi_{3}^{\mu}(0) | N_{1/2^{+}}(\vec{p},s) \rangle\langle N_{1/2^{+}}(\vec{p},s) |\bar{\chi}_{i}(0) | 0 \rangle \cr
&=& \bar{\lambda}_{N_{1/2^{+}}}\
(\alpha_{N_{1/2^{+}}} p_{N_{1/2^{+}}}^{\mu} + \beta_{N_{1/2^{+}}}\gamma^{\mu})\ \gamma_5 {(\gamma \cdot p_{N_{1/2^{+}}} + M_{N_{1/2^{+}}})\over 2E_{N_{1/2^{+}}}} \ .
\end{eqnarray}


At $\vec{p}=\vec{0}$, the positive parity states for $\mu=1$ propagate
in the real part of the $(1,2)$ and $(2,1)$ spinor elements of the
correlation function.
For $\mu=2$, the positive parity states propagate in the imaginary
part of the $(1,2)$ and $(2,1)$ elements, with a relative minus sign.
For $\mu=3$ the positive parity states propagate in the real part
of the $(1,1)$ and $(2,2)$ elements, with a relative minus sign,
and for $\mu=0$ they propagate in the real part of the $(1,3)$
and $(2,4)$ elements.

Similarly, the odd parity contribution to the correlation function is:
\begin{eqnarray}
& &\sum_{s} \langle 0 | \chi_{3}^{\mu}(0) | N_{1/2^{-}}(\vec{p},s) \rangle\langle N_{1/2^{-}}(\vec{p},s) |\bar{\chi}_{i}(0) | 0 \rangle \cr
&=& \bar{\lambda}_{N_{1/2^{-}}}\
(\alpha_{N_{1/2^{-}}} p_{N_{1/2^{-}}}^{\mu} + \beta_{N_{1/2^{-}}}\gamma^{\mu})\gamma_{5} {(\gamma \cdot p_{N_{1/2^{-}}} - M_{N_{1/2^{-}}})\over 2E_{N_{1/2^{-}}}} \ .
\end{eqnarray}
Combining the even and odd parity contributions, we obtain for the
``$3i$'' correlation function:
\begin{eqnarray}
\mathcal{G}_{3i}^{\mu}(t,{\vec p}) &=&  { \sum_{B^{+}} } e^{-E_{B^{+}}t}
\bar{\lambda}_{N_{1/2^{+}}}\
(\alpha_{N_{1/2^{+}}} p_{N_{1/2^{+}}}^{\mu} + \beta_{N_{1/2^{+}}}\gamma^{\mu})\
\gamma_5 {(\gamma \cdot p_{N_{1/2^{+}}} + M_{N_{1/2^{+}}})\over 2E_{N_{1/2^{+}}}} \cr
&+& { \sum_{B^{-}} } e^{-E_{B^{-}}t}
 \bar{\lambda}_{N_{1/2^{-}}}\
(\alpha_{N_{1/2^{-}}} p_{N_{1/2^{-}}}^{\mu} + \beta_{N_{1/2^{-}}}\gamma^{\mu})\gamma_{5} {(\gamma \cdot p_{N_{1/2^{-}}} - M_{N_{1/2^{-}}})\over 2E_{N_{1/2^{-}}}} \ .
\end{eqnarray}
Using the appropriate terms in Eqs.~(\ref{eq:even_matrix_element})
and (\ref{eq:lorentz_pseudovector_elements}) in
Eq.~\eqref{eq:2-pt.cmatrix-xmu}, the ``$i3$'' correlation function can
be written:
\begin{eqnarray}
\mathcal{G}_{i3}^{\mu}(t,{\vec p}) &=&  - { \sum_{B^{+}} } e^{-E_{B^{+}}t}\
\lambda_{N_{1/2^{+}}}\
{(\gamma \cdot p_{N_{1/2^{+}}} + M_{N_{1/2^{+}}})\over 2E_{N_{1/2^{+}}}}\
\gamma_{5}(\bar{\alpha}_{N_{1/2^{+}}}p_{N_{1/2^{+}}}^{\mu} + \bar{\beta}_{N_{1/2^{+}}} \gamma^{\mu}) \cr
%
&+& { \sum_{B^{-}} } e^{-E_{B^{-}}t}\
\lambda_{N^{1/2^{-}}}\ 
\gamma_{5}{(\gamma \cdot p_{N^{1/2^{-}}} + M_{N^{1/2^{-}}})\over 2E_{N_{1/2^{-}}}}\
(\bar{\alpha}_{N^{1/2^{-}}}p_{N^{1/2^{-}}}^{\mu} + \bar{\beta}_{N^{1/2^{-}}} \gamma^{\mu})\ .
\end{eqnarray}
These functions can then be used to relate the appropriate elements
of the correlation function to a particular parity.
To improve our statistics we will average the correlation functions
over the spatial components of $\chi^{\mu}_{3}$.

\end{widetext}

\section{ Results }
\label{sec:res}

We begin our analysis of the spectrum by considering the $2\times 2$
correlation matrices with $\chi_1$ and $\chi_3^\mu$, and with $\chi_2$
and $\chi_3^\mu$.  The $2\times 2$ correlation matrix with $\chi_1$ and
$\chi_2$ has previously been explored in
Ref.~\cite{Melnitchouk:2002eg}.  To choose a time slice at which to
invert the correlation matrix, we determine the earliest plateau
available to the individual interpolators.  Our initial time is taken to
be one time slice earlier than this.  If this analysis fails, we
invert the correlation matrix at one time slice earlier.  This
algorithm is discussed in more detail in Ref.~\cite{Lasscock:2005tt}.
Throughout our correlation matrix analysis we only consider a shift
of one time slice from this inversion time.

Proceeding with the $2\times 2$ correlation matrices, the eigenvectors
for the projection of the $\chi_1$, $\chi_3^\mu$ correlation matrix are
obtained from an analysis at time slice $t=14$, which is two steps
back from the onset of the plateau of the effective mass extracted
with $\chi_3^\mu$, and six time slices after the source.  For the
$\chi_2$, $\chi_3^\mu$ correlation matrix, the onset of the plateau in the
effective mass extracted with the $\chi_2$ interpolator is at $t=11$.
Eigenvectors for the projection of the $\chi_2$, $\chi_3^\mu$ correlation
matrix are therefore obtained from an analysis at $t=10$.

\begin{figure*}[tbp]
\begin{center}
\begin{tabular}{lr}
\includegraphics[height=0.5\hsize,angle=90]{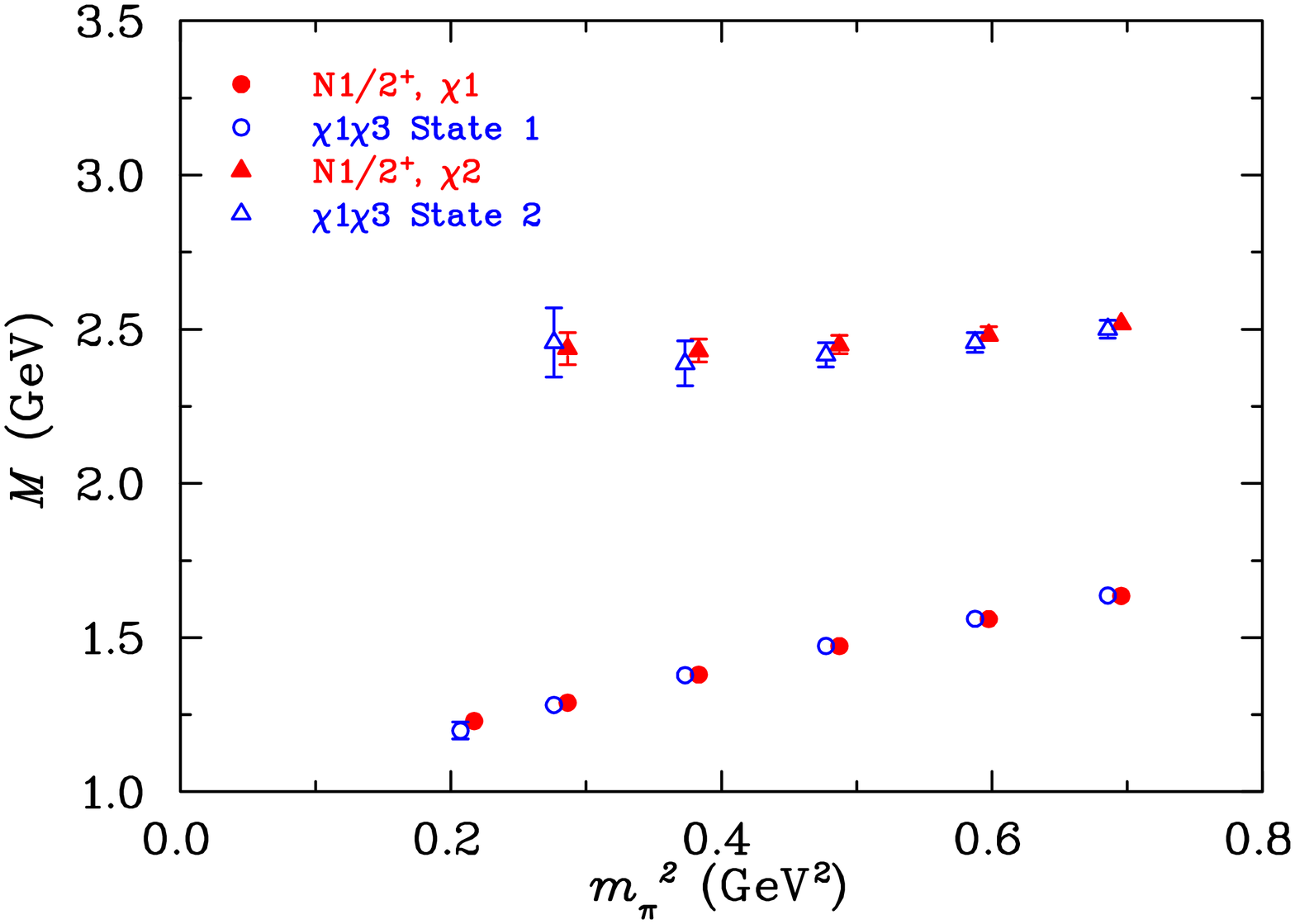} &
\includegraphics[height=0.5\hsize,angle=90]{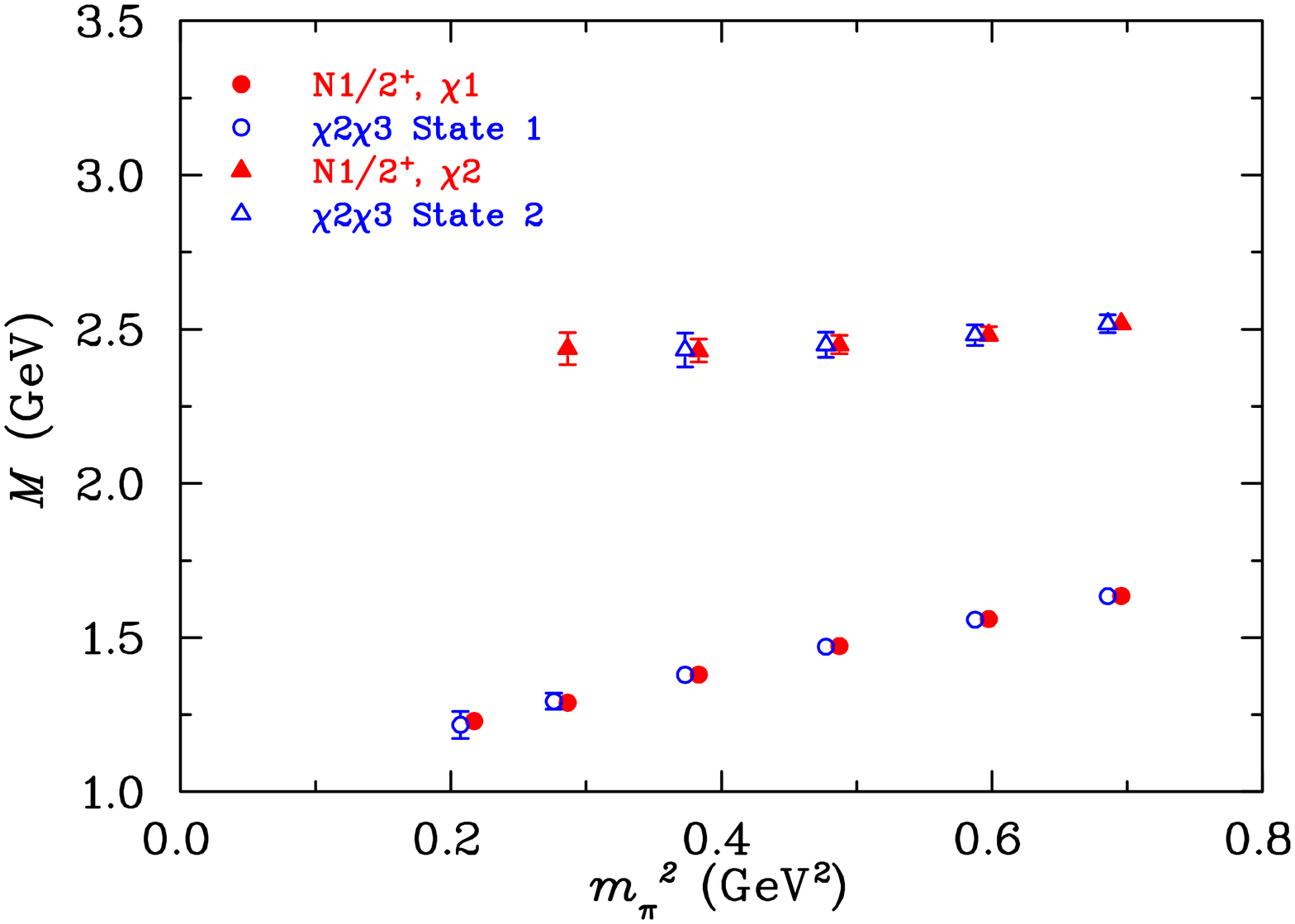}  
\end{tabular}
\end{center}
\caption{\label{fig:Masses.2x2}
  Masses extracted with a $2\times 2$ correlation matrix with
  $\chi_{1}$ and $\chi_{3}^\mu$ (left), and $\chi_{2}$ and $\chi_{3}^\mu$
  (right).
  For comparison the masses extracted with the $\chi_{1}$ and $\chi_{2}$
  interpolators are also shown.  The data correspond to $\mpi\simeq 830$
  (rightmost points), 770, 700, 616, 530 and 460~MeV (leftmost points).}
\end{figure*}

The masses of the ground and excited states for each correlation matrix
are shown in Fig.~\ref{fig:Masses.2x2}, along with the masses extracted
with the $\chi_{1}$ and $\chi_{2}$ interpolators individually.
In each case we find that the mass of the ground state extracted with
the correlation matrix analysis is in excellent agreement with the mass
of the state extracted with the $\chi_{1}$ interpolator.
Furthermore, the mass of the excited state extracted with the
correlation matrix agrees well with the mass extracted with the
$\chi_{2}$ interpolator.
The previous CSSM study \cite{Melnitchouk:2002eg} showed that the
$\chi_{1}$ interpolator is largely orthogonal to $\chi_{2}$.
The present calculation shows, therefore, that the $\chi_{3}^\mu$
interpolator has a significant overlap with the states accessed
by both $\chi_{1}$ and $\chi_{2}$.

\begin{widetext}
With this information in mind we need to determine if the spin-1/2
projected $\chi_3^\mu$ interpolator is a simple linear combination of
$\chi_1$ and $\chi_2$. Using the Fierz identity:
\begin{eqnarray}
\delta_{\alpha\alpha'}\delta_{\beta\beta'} = \frac{1}{4} \sum_{J} (\Gamma_{J})_{\alpha\beta'}(\Gamma_{J}^{-1})_{\beta\alpha'}, \ 
\end{eqnarray}
where $\Gamma$ is one of the matrices in the Dirac representation
$\{ 1, \gamma_{5}, \gamma_{\mu}, \gamma_{\mu}\gamma_{5},
    \sigma_{\mu\nu}|_{\mu >\nu}\}$\ ,
one can show that:
\begin{eqnarray}
\label{eq:chi3_fierzed}
\chi^{\mu}_{3}
&=& \frac{1}{4} \epsilon^{abc}(u^{a T}C\gamma^{\sigma}u^{b})\gamma_{\sigma}\gamma^{\mu}d^{c}\ 
 -  \frac{1}{8} \epsilon^{abc}(u^{a T}C\sigma^{\sigma\rho}u^{b})\sigma_{\sigma\rho}\gamma^{\mu}d^{c}\ .
\end{eqnarray}
Acting on $\chi_3^\mu$ with the spin-1/2 projector from
Ref.~\cite{Zanotti:2003fx}, we can write:
\begin{eqnarray}
P^{1/2}_{\mu\nu}\chi^{\nu}_{3}
&=& \frac{1}{3}\gamma_{\mu}\gamma_{\nu}\chi_{3}^{\nu}
 +\ \frac{1}{3p^{2}}(\gamma\cdot p \gamma_{\mu}p_{\nu}
		   + \gamma_{\nu}p_{\mu}\gamma\cdot p) \chi_{3}^{\nu}
\end{eqnarray}
Expanding the combination $\gamma_{\nu}\chi^{\nu}_{3}$ using
Eq.~\eqref{eq:chi3_fierzed}, we obtain:
\begin{eqnarray}
\gamma_{\nu}\chi^{\nu}_{3}
&=& \frac{1}{4} \epsilon^{abc}(u^{a T}C\gamma^{\sigma}u^{b})\
                \gamma_{\nu}\gamma_{\sigma}\gamma^{\nu}d^{c} 
  - \frac{1}{8} \epsilon^{abc}(u^{a T}C\sigma^{\sigma\rho}u^{b})\
                \gamma_{\nu}\sigma_{\sigma\rho}\gamma^{\nu}d^{c}\nonumber\\
&=& -\frac{1}{2}\epsilon^{abc}(u^{a T}C\gamma^{\sigma}u^{b})
		\gamma_{\sigma}d^{c}				\nonumber\\
&=& \frac{1}{2}\gamma_{5} \chi_{{\rm SR}}\ ,
\end{eqnarray}
where we identify $\chi_{{\rm SR}}$ from Ref.~\cite{Leinweber:1994nm}
as the linear combination
$2(\chi_{2} - \chi_{1})$.
Substituting this result in Eq.~\eqref{eq:chi3_fierzed}, we find:
\begin{eqnarray}
  P^{1/2}_{\mu\nu}\chi^{\nu}_{3} &=& \frac{1}{6}\gamma_{\mu}\gamma_{5}\chi_{{\rm SR}} +\
   \frac{1}{3p^{2}}(\gamma\cdot p \gamma_{\mu}p_{\nu} + 2p_{\mu}p_{\nu})\chi^{\nu}_{3}\
 - \frac{1}{6p^{2}}( \gamma\cdot p p_{\mu} \gamma_{5} \chi_{{\rm SR}} )\ .
\end{eqnarray}
At zero momentum the spin-1/2 projected $\chi_3^\mu$ can then be written:
\begin{eqnarray}
  P^{1/2}_{k\nu}\chi^{\nu}_{3} &=& \frac{1}{3}\gamma_{5}\gamma_{k}(\chi_{1} - \chi_{2}) + \frac{1}{3}\gamma_{0}\gamma_{k}\chi_{3}^{0}
\end{eqnarray}
\end{widetext}
Since we consider the spatial components of the spin projected
correlation function, $\mu=1$--3, the only new information in the
$\chi_3^\mu$ interpolator is from $\chi^{0}_{3}$, i.e. the time
component of the $\chi_3^\mu$ interpolator used by Br\"ommel
et~al. \cite{Brommel:2003jm}. Thus our analysis serves as an
independent check of Ref.~\cite{Brommel:2003jm}. 

The Fierz transformation of the $\chi_3^\mu$ interpolator allows us to
verify that this has a strong overlap with both the $\chi_{1}$ and
$\chi_{2}$ fields.
The question remains whether there is sufficient additional information
in the $\chi_3^\mu$ interpolator to extract a second excited state.
We proceed, therefore, with a $3\times 3$ correlation matrix analysis
with $\chi_{1}$, $\chi_{2}$ and $\chi_{3}^\mu$.
Eigenvectors for the projection of the correlation matrix are obtained
from an analysis at $t=10$, one step back from the onset of the plateau
in the effective mass extracted with the $\chi_2$ interpolator. 

\begin{figure}[tbp]
\begin{center}
\includegraphics[height=1.0\hsize,angle=90]{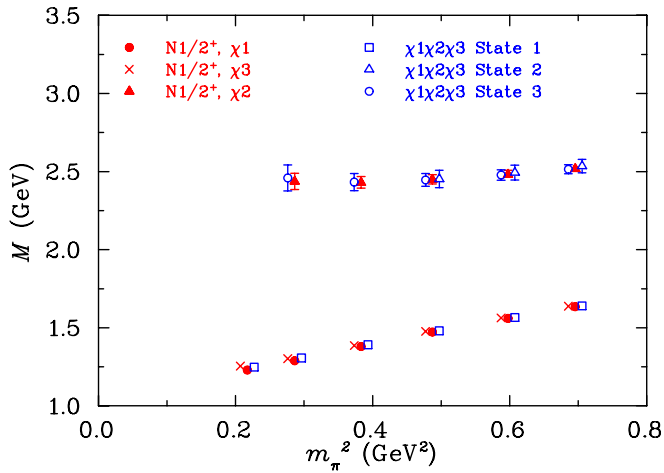} 
\end{center}
\caption{\label{fig:Masses.X123}
	As in Fig.~\ref{fig:Masses.2x2}, but for the $3\times 3$
	correlation matrix with the $\chi_{1}$, $\chi_{2}$ and
	$\chi_{3}^\mu$ interpolators.}
\end{figure}

The extracted masses are shown in Fig.~\ref{fig:Masses.X123}, along
with the masses extracted with the $\chi_1$, $\chi_2$ and $\chi_3^\mu$
interpolators individually.  All three determinations of the ground
state mass are found to be in excellent agreement, as are the masses
of the excited state extracted with the correlation matrix, and the
$\chi_2$ interpolator.  At the two smaller quark masses we only have
sufficent statistics to fit two of the three diagonal elements of the
projected correlation matrix. The masses extracted with the $\chi_1$ and 
$\chi_2$ interpolators individually are shown in Table.~\ref{tab:masses}.
\input{masses.tbl}

Our results show no evidence of a Roper-like even parity excited state,
which suggests that the couplings of the interpolators used in this
study to such a state must be either small or zero.  
On the other hand,
using Bayesian techniques the analyses in
Refs.~\cite{Sasaki:2005ap,Mathur:2003zf} do report a low-lying state.
If the Roper resonance exists in quenched QCD near its experimental 
value, then
it would appear that our smeared sources have unfortunately poor
overlap with this state. 
However we note that Br\"ommel et~al. \cite{Brommel:2003jm}, Burch et~al.
\cite{Burch:2006cc}, and Basak et~al. \cite{Basak:2006ww} all report
results similar to ours with different interpolating fields.  It is
difficult to understand, therefore, if this state does actually exist,
why it would not be seen in any of these analyses.

\begin{figure*}
\includegraphics[width=0.75\hsize]{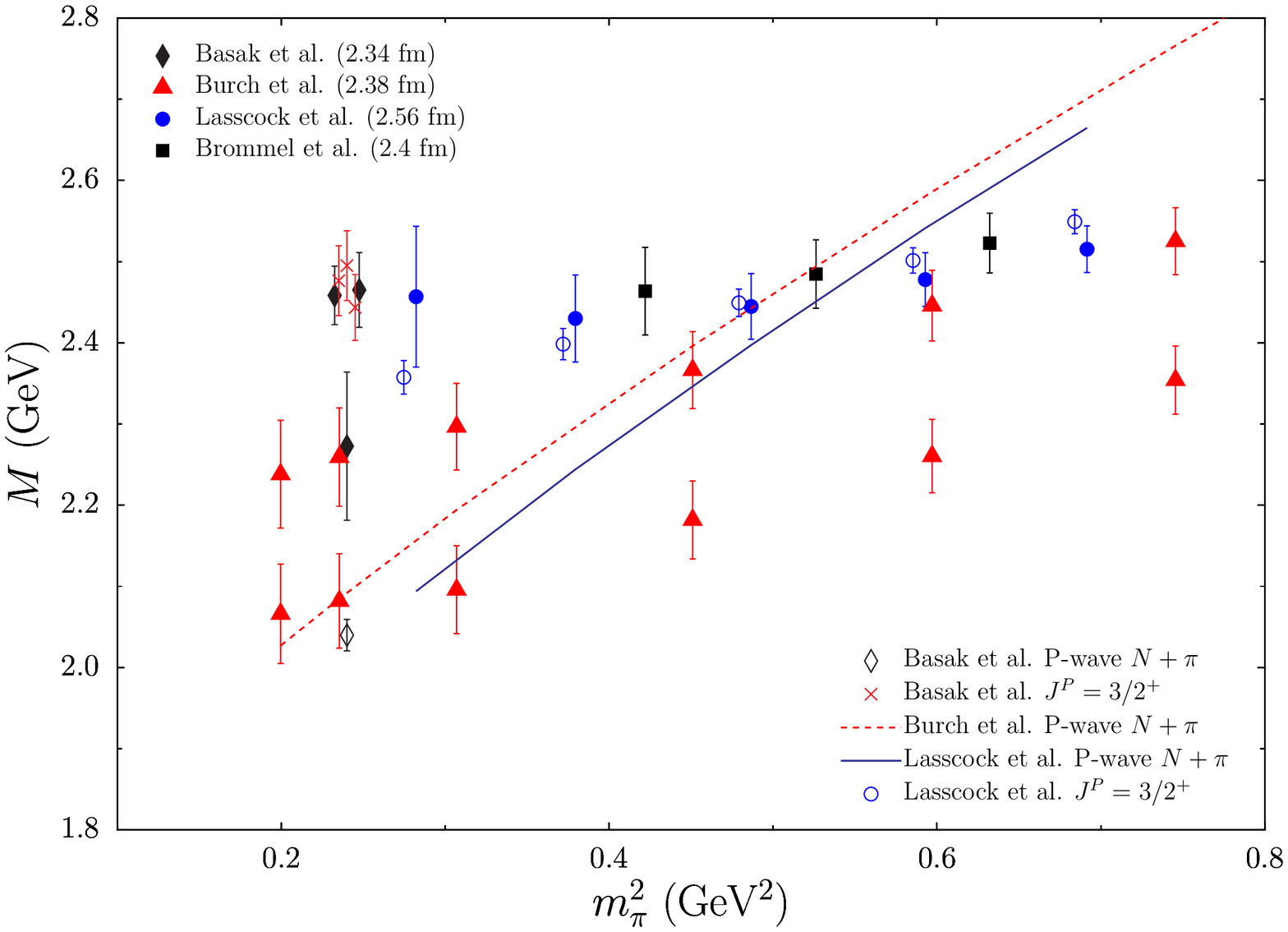}
\caption{\label{fig:zoomed} A summary of existing lattice calculations
  of the nucleon spectrum based on correlation matrix techniques.  The
  data points with degenerate masses from Basak et~al.
  \cite{Basak:2006ww} have been
  displaced horizontally for clarity.  P-wave two particle energies
  are illustrated by the two lines.  The difference in the energies
  reflects the difference in the lattice volumes.  }
\end{figure*}

In Fig.~\ref{fig:zoomed} we enlarge a portion of
Fig.~\ref{fig:compilation}, adding the energies of the non-interacting
P-wave $N+\pi$ for each study, and the mass of the even parity
spin-3/2 nucleon calculated here and in Ref.~\cite{Basak:2006ww}.  We
show the mass of the spin-3/2 state as a guide because it has a mass
of $1720$ MeV, similar to the mass of the $N'$. Further, at the larger
quark masses it is reasonable to expect that the hyperfine splitting
between the spin-1/2 and spin-3/2 states should be small.  Recall that
the energy of the P-wave $N+\pi$ threshold state will be larger than
the attractive scattering state mass due to finite volume effects and
that a stable state is likely to appear at sufficiently heavy quark masses
\cite{Lasscock:2005tt,Lasscock:2005kx}.  Note that for the data from
Ref.~\cite{Basak:2006ww} labeled as spin-3/2, only one of the three
degenerate masses corresponds to a spin-3/2 state, with the others
corresponding to higher spin states.

At the three largest quark masses, the mass of the excited state
calculated in this study is consistent with the highest-lying
excited state calculated by Burch et~al.\ and with the spin-3/2 state.
We also reproduce the mass of the excited state calculated by Br\"ommel
et~al.\ .
%
As we approach smaller quark masses,  the mass of the observed excited
state becomes larger than the energy of the non-interacting P-wave
$N+\pi$ state.  
%
In this case the reported masses will include some contamination from
the $N\pi$ threshold, although the spectral strength may be relatively
weak.  Even with this low-mass contamination there is no evidence of a
state approaching $1440$ MeV.

\section{Conclusion}
\label{sec:conc}

In this study we have attempted to extract the mass of the first even
parity excitation of the nucleon in quenched lattice QCD, using a basis
of three nucleon interpolating fields.  We have demonstrated that the
ground and second even parity excited states of the nucleon can be
extracted from a $2\times 2$ correlation matrix, with either the
$\chi_1$ and $\chi_3^\mu$, or $\chi_2$ and $\chi_3^\mu$ interpolating
fields.  With the use of the Fierz identity, we showed that the
spin-1/2 projected $\chi_3^\mu$ interpolator does indeed have a large
overlap with the $\chi_1$ and $\chi_2$ interpolators,
and with $\chi^{0}_{3}$.

Extending the analysis to a $3\times 3$ correlation matrix, we found no
evidence that our interpolators overlap with the Roper resonance.
Our results are in accord with other correlation matrix based analyses
using similar interpolating field constructs.  The absence of a low-lying
excitation in any of these analyses raises the question of how this state
is seen in Bayesian analyses.  Since the multiple-operator correlation
matrix approach is an established method for extracting excited state
masses, 
further careful examination of the Bayesian analyses would seem appropriate.

By comparison with the mass of the even parity spin-3/2 state at large
quark masses we identify the excited state extracted in our
correlation matrix as the $N'(1710)$.  However we recognize that
approaching the chiral regime the level ordering of the $N'$ and
the P-wave $N+\pi$ (and P-wave $N+\eta'$ in quenched QCD) is
reversed. This ambiguity makes an analysis to discriminate scattering 
states central to future calculations of the spectrum at light 
quark masses.

On a final note we emphasize the need to bring these advanced analysis
techniques to bare on dynamical fermion configurations. There is now
ample evidence that this is likely to be the key missing ingredient in
creating and observing the Roper resonance in lattice QCD simulations.

\section*{Acknowledgements}
We thank the Australian Partnership for Advanced Computing (APAC) and
the South Australian Partnership for Advanced Computing (SAPAC) for
generous grants of supercomputer time which have enabled this project.
This work was supported by the Australian Research Council.  W.~M. is
supported by Jefferson Science Associates, LLC under U.S. DOE Contract
No. DE-AC05-06OR23177. J.Z. is supported by PPARC grant PP/D000238/1.

\bibliographystyle{apsrev}
\bibliography{roper}

\end{document}

%% file: masses.tbl
\begin{table}[htbp]
\caption{\label{tab:masses} The mass of the nucleon extracted with $\chi_1$ and the even-parity spin-1/2 state excited 
state extracted with $\chi_2$. For comparison we show the mass of the even-parity spin-3/2 state extracted with 
$\chi_3$.}
\begin{ruledtabular}
\begin{tabular}{cccc}
$a m_\pi$ & $m_{N}$ & $m_{N'}$ & $m^{3/2^{+}}_{N}$ \cr
\hline
0.541(1) & 1.061(3)  & 1.633(14) & 1.655(9)  \cr
0.501(1) & 1.012(4)  & 1.609(18) & 1.624(10) \cr
0.454(1) & 0.955(5)  & 1.589(21) & 1.590(11) \cr
0.401(1) & 0.895(6)  & 1.576(25) & 1.557(13) \cr
0.347(2) & 0.836(9)  & 1.581(34) & 1.531(13) \cr
0.302(2) & 0.790(12) & ...       & 1.515(16) \cr
0.241(4) & 0.755(13) & ...       & 1.515(19) \cr
0.197(5) & 0.745(14) & ...       & 1.545(26) \cr
\end{tabular}
\end{ruledtabular}
\end{table}